# Graphene to Graphane: Novel Electrochemical Conversion

Kevin M. Daniels[1,a], B.Daas[1], R. Zhang[2], I. Chowdhury[1], A. Obe[2], J. Weidner[2], C. Williams[2], T.S. Sudarshan[1], MVS Chandrashekhar[1]

*(1) Department of Electrical Engineering, University of South Carolina, Columbia, SC, 29208, USA*
*(2) Department of Chemical Engineering, University of South Carolina, Columbia, SC, 29208, USA*
*[a]danielkm@email.sc.edu*

**Abstract**

A novel electrochemical means to generate atomic hydrogen, simplifying the synthesis and controllability of graphane formation on graphene is presented. High quality, vacuum grown epitaxial graphene (EG) was used as starting material for graphane conversion. A home-built electrochemical cell with Pt wire and exposed graphene as the anode and cathode, respectively, was used to attract H+ ions to react with the exposed graphene. Cyclic voltammetry of the cell revealed the potential of the conversion reaction as well as oxidation and reduction peaks, suggesting the possibility of electrochemically reversible hydrogenation. A sharp increase in D peak in the Raman spectra of EG, increase of D/G ratio, introduction of a peak at ~2930 $cm^{-1}$ and respective peak shifts as well as a sharp increase in resistance showed the successful hydrogenation of EG. This conversion was distinguished from lattice damage by thermal reversal back to graphene at 1000°C.

## Introduction

Epitaxial graphene has sparked much interest in the materials and device community, boasting such properties as high electrical conductivity, high thermal conductivity and high tensile strength due to the C-C bond, high room temperature mobility, quasi-ballistic transport of carriers, low noise [1-4] and ability to be produced on large areas on commercial SiC substrates [19]. Formation of epitaxial graphene is done by thermal desorption of Si from SiC in a vacuum [4] or in argon [5,17]. One of the unique properties of graphene is its zero-bandgap, making it a semi-metal. This zero-gap precludes the fabrication of traditional semiconductor devices due to excessive band-to-band leakage.

However, when hydrogen reacts with the $\pi$-bonds (Fig. 1) of graphene, the delocalized $\pi$-electron becomes localized at the C-H bond, decreasing the conductivity, changing the bond hybridization from $sp^2$ to $sp^3$ and increasing the C-C bond length ~7% from 1.42A to 1.52A [6]. This hydride of graphene is called graphAne. The electron localization at the $sp^3$ bond leads to a decrease in conductivity, and opens up a bandgap varying from 0-3.5eV depending on the degree of hydrogenation [6]. This enables new applications in bandgap engineered electronics using carbon-based materials. Furthermore, graphane is as thermodynamically stable as comparable hydrocarbons, more stable than metal hydrides and more stable than graphene by ~0.15eV [6]. This along with its large hydrogen storage capacity 7.7 wt%, which exceeds the Department of Energy (DOE) 2010 target of 6% also makes it a promising candidate for hydrogen storage [6].

To form graphane, graphene must react with hydrogen. While bulk graphite has been observed to be chemically inert, graphene has shown enhanced reactivity with atoms such as fluorine [12] and most importantly, hydrogen [6]. Sharma et. al. demonstrated graphene's enhanced chemical reactivity by functionalizing it with 4-nitrobenzene diazonium tetrafluoroborate. This functionalization was done on a single monolayer, bi-layer and bulk graphite, which showed that a single monolayer of graphene is not only more reactive than bulk graphite, but up to ten times more reactive than even a bi-layer of graphene. The extent of the reactivity of graphene was quantified by an increase in Raman D peak resulting in a higher D/G ratio after functionalization [7]. This point will be addressed in greater detail in the context of the measurements in this paper.

The enhanced reactivity was attributed to substrate induced electron transfer between the substrate and graphene, leading to a shift in the energetics of the graphene layer compared to bulk graphite, and consequently affecting the reactivity. Since the electrostatic screening length in graphene is ~1 monolayer (ML), only the first layer exhibits this enhanced reactivity [7,20]. This first monolayer experiences potential fluctuations from external impurity charges while subsequent graphene layers are electrically neutral, exhibiting the reactivity of bulk graphite. Other potential factors influencing the reactivity were also pointed out; a) The presence of wrinkles on the graphene surface [8] and b) electron-hole puddles from surface impurities leading to local enhancement of chemical reactivity. These were, however, considered to be minor contributions. The difficulty with reacting graphene to form graphane is the need for atomic hydrogen, as hydrogen gas $H_2$ is unreactive. Techniques implemented by other groups involve *in situ* development of atomic hydrogen by plasma-assistance [8] or hot filaments [14], as the H-H bond in hydrogen gas requires high energy/temperature to break [8].

With the limitations of current techniques to form graphane presented, an alternative, electrochemical means to synthesize graphane is demonstrated in this paper. Electrochemistry offers the most controlled route to the systematic synthesis of graphane, as the extent of the hydrogenation of graphene can be precisely controlled by changing the current level (or voltage) and time. Such control is not as easily achievable using the techniques described above. Furthermore, through electrochemistry, reactions can be conducted at ambient conditions, as opposed to the harsh environments in other techniques. The convenience and controllability of electrochemical hydrogenation of graphene therefore provides a more realistic approach for a tunable bandgap in graphene/graphane. In this paper, atomic hydrogen is simply generated through an acidic electrolyte ($H_2SO_4$-sulfuric acid), giving free $H^+$ ions that can then be easily attracted to an appropriately biased negative graphene cathode for reaction to form graphane. This technique is also compatible with conventional semiconductor fabrication technologies. For example, the sulfuric acid electrochemistry can be incorporated directly into lithographically defined patterns with e-beam photoresist hydrogen silsesquioxane (HSQ), which is compatible with dilute acids. Performing lithography and then the conversion in the manner described in the manuscript it would be possible to form graphane and graphene channels on a single sample. Other acids are also compatible with organic photoresists, for example dilute HF.

**Experimental Details**

6H SiC semi-insulating, on-axis wafers were obtained from II-IV, Inc. for this study. The Si-face was chemical-mechanical polished (CMP) and the C-face was optically polished. 10 x 10 $mm^2$ samples were diced from these wafers and thoroughly cleaned using a standard RCA clean (TCE, acetone, methanol) and HF to remove any naïve oxide. Growth of epitaxial graphene (EG) was done on both Si and C-faces by thermal decomposition of a SiC substrate *in vacuo* [16], $<10^{-5}$ Torr, using an RF furnace. Temperatures for growth ranged form 1250-1450°C, with quality of EG growth verified with atomic force microscopy (AFM) and Raman, showing D/G ratios < 0.2, demonstrating the high quality of the starting material [10]. EG layer thicknesses were estimated by x-ray photoelectron spectroscopy, Raman spectroscopy [11] and infrared transmission measurements [13].

Atomic hydrogen was generated using a home-built electrochemical setup, as shown in Fig. 2, with current applied though a 10% $H_2SO_4$ acid solution. A 99.6% Pt wire and exposed EG (approximately a 4mm diameter circular area) were used as the anode and cathode, respectively. With this setup, H+ ions in the $H_2SO_4$ electrolyte are attracted to the exposed graphene. Cyclic voltammetry (CV), with an $Hg_2SO_4$ reference (0.67V vs. NHE) was used to determine the graphane conversion potential. Oxidation occurs at the Pt anode during this process, and has been investigated and discussed in greater detail elsewhere [21,23]. Thus, in this investigation, we limit ourselves to the reduction of the graphene cathode.

Using the determined potential, future conversions to graphane were performed until the current through the graphane decreased <10 nA, typically after ~1 hour. This conversion was confirmed using graphane's known Raman peaks [9] and an increase in 2 terminal resistance [6], both of which are compared to the pre-converted sample and a non-converted area on the graphene sample after conversion.

Raman spectroscopy was performed using a micro-Raman setup with laser excitation wavelength at 632nm with a spot size of ~2 μm. The Raman system was calibrated using the known Si peak at 520.7 $cm^{-1}$. Reference blank substrate spectra were scaled appropriately and subtracted from the EG spectra to show only the graphene and graphane peaks [8,10]. All the spectra shown in this paper are difference Raman spectra obtained in this manner. Raman was used as an indication of hydrogenation and reversal by the behavior of the D, G and 2D peaks, which corresponds to the disorder-induced peak, in-plane vibrations and double resonant [11], respectively, as well as a C-H bond peak introduced at ~2930 $cm^{-1}$ [8]. The D peak is expected to increase as a function of hydrogenation [7, 8] as well as the introduction of a D' shoulder on the G peak [8], which supports the claim of $sp^3$ hybridization after hydrogenation [8].

The 2 terminal resistance measurements were performed with one probe on the same external point as a reference on each sample and the other probe within and around the conversion area. The final confirmation of graphane is thermal reversal back to graphene which is done through a 1000°C anneal [8] *in vacuo,* where hydrogen desorbs from the graphene, restoring it to its pre-hydrogenated state. This thermal process was done for 4 hours and 50 hours to understand the time dependence of this reversal. A control sample at 1000C showed no measurable EG growth, confirming that the recovery is indeed due to hydrogen desorption rather than growth of high quality EG.

Electrical measurements were made possible by the conversion to graphane on EG grown on unintentionally doped n- epitaxial SiC ($3x10^{15}$ $cm^{-2}$) on n+ SiC substrates (Fig. 9). The n- SiC epilayer was grown by CVD in a vertical hot wall chimney style reactor, using propane and dichlorosilane precursors with growth temperature at 1550°C and pressure 80 Torr at a growth rate of 30um/hr .[16] The sample area was 8mm x 8mm with a converted 4mm diameter circle in the middle, which allowed for numerous 300µm diameter Schottky diodes to be fabricated using a shadow mask on the graphene and graphane areas. 1200Å Ni was used as the top metal contact and as a hard mask for mesa-isolation. $O_2$ plasma reactive-ion etching (RIE), was used to electrically isolate each Schottky diode. Raman on the non-mesa parts of the sample confirmed successful removal of non-device graphene/graphane and device isolation. With conductive graphene on the backside forming an ohmic large area contact on the highly doped substrate, the metal chuck was used as the backside contact during the measurements.

**Results and Discussion**

Figure 3 shows a cyclic voltammogram of the EG/10% $H_2SO_4$/Pt electrochemical cell (Fig. 2) at a sweep rate of 20mV/s. The EG working electrode's potential, was cycled to determine the potential at which hydrogen incorporates into the EG. Using this cyclic voltammogram, it was determined that the conversion occurs 0.2V below the evolution of hydrogen by taking the middle point between the adsorption and desorption peaks of hydrogen (15mV and 258mV), similar to that seen with other metal hydrides [21, 23]. The presence of both adsorption and desorption peaks of hydrogen are indicative of a possibly electrochemically reversible system, allowing for hydrogen to be removed simply by switching the polarity of the system. This opens up new possibilities in the field of efficient hydrogen storage, as hydrogen can directly be incorporated into the storage medium at ~1V anode/cathode potential vs. ~1.5-1.8V for electrolytic hydrogen gas [18] which must then be reacted to form a stable hydride, requiring more energy. In other words, the direct electrochemical hydrogen loading/unloading in this system may make hydrogen storage ~2x more efficient than currently possible. While this reversibility is surprising, it can be attributed to charge transfer at the SiC/EG interface[7], leading to non-neutral EG layers at that interface, as was observed with exfoliated graphene on $SiO_2$ [7].

With the conversion potential obtained from cyclic voltammetry, EG-hydrogenation reactions were carried out on Si-face and C-face EG samples. A typical Current (I) vs. time (t) curve of such a reaction is shown in Figure 4, with reactions cut off after 1 hour, as discussed above. Based on the total integrated charge of the I vs. t curve, ~1ML of hydrogen was incorporated into 5ML of graphene. We note that there was no observable influence of substrate polarity and EG-layer thickness on the hydrogen loading of EG. In other words, 1ML of hydrogen was incorporated regardless of the starting EG layers. While we cannot conclusively ascertain where this hydrogen has been incorporated, we speculate that it occurs at the SiC-substrate/EG interface, where there is potentially enhanced reactivity [7]. This is also in agreement with high temperature hydrogenation experiments with EG [14], where hydrogen reacts first at the interface, rather than at the EG surface. This may be explained by charge transfer at the substrate/graphene interface, similar to that observed by Sharma et. al [7] on exfoliated graphene, although further investigation is required to elucidate this mechanism further.

AFM images of Si and C-face EG before conversion (Fig. 5a and b) show the steps and giraffe stripes, respectively, which are common in EG growth on these faces [19]. Post-conversion morphologies show a slight increase in root-mean square (RMS) roughness, from 0.6 to 1.0nm on Si-face and 2.8 to 2.9nm on C-face. What is interesting about the Si and C-face graphane surfaces (Fig. 5c and Fig. 5d) is the addition of raised streaks and patches on the Si and C-face respectively. The streaks in the Si-face in particular follow the step direction. On the C-face, the wrinkles on the surface associated with the giraffe stripes disappear after conversion. A two-probe resistance measurement also showed a significant increase in resistance in these areas. With this Si-face sample in particular, graphene resistance was approximately 9kΩ. After conversion, this resistance was found to be

>40MΩ. The morphology changes as seen using AFM and the increase in resistance are two indicators of a successful conversion.

The Raman peaks of graphene and graphane were observed with micro-Raman (Fig. 6). As expected, there was a sharp increase in the D peak [6] corresponding to the functionalization of the graphene. A red shift in the D peak, from 1350cm$^{-1}$ to 1330cm$^{-1}$, was also observed, likely caused by the formation of sp$^3$ bonds [8, 10]. The G-peak FWHM also broadens with hydrogenation further supporting the sp$^3$ hybridization of the graphene [8]. Another peak at ~2930cm$^{-1}$ was an indication of C-H bonds [8]. A fluorescence background, along with increased SiC substrate signal was also observed in the working area, suggesting the presence of a bandgap in the material. Such a fluorescence background was not observed in the starting EG.

As shown in Fig. 7, there was also a gradient associated with the conversion, where, according to Raman, the center showed significant hydrogenation when compared to other parts within the converted area. This gradient showed significant red shifting of the D peak, 1349 to 1328 cm$^{-1}$, from just outside the conversion area to the center, as well as a considerable red shift in the 2D peak from 2700 to 2626 cm$^{-1}$. The intensity of the D peak also increases, and as a result, the D/G ratio also changes within the gradient, with the ratio changing from 0.06 outside the conversion area to 1.35 at the center. This gradient in functionalization could be caused by the graphene becoming electrically decoupled from the substrate during the hydrogenation process [14].

While these Raman signatures are compelling evidence for hydrogen incorporation, they can also be interpreted on the basis of lattice damage, which would provide similar changes in the Raman spectrum [10]. Therefore, to distinguish this conversion from lattice damage, a reversal of graphane back to EG was performed by annealing in vacuum for 4 hours and 50 hours at 1000°C. The Raman spectra of the area (Fig.6) after reversal clearly shows the disappearance of the C-H peak at ~2930cm$^{-1}$ indicating desorption of hydrogen in the material. However, after the 4 hour anneal, a D/G ratio of ~1, the continued presence of fluorescence, and a sufficiently high resistance of 100kΩ indicated that some hydrogen still remained. With hydrogen still present, a 50 hour anneal was done to ensure that all hydrogen was removed from the graphene. After the 50 hour anneal the D peak shifted back to pre-conversion state at 1340 cm$^{-1}$ and the fluorescence background was no longer present. While the D and 2D peaks shifted back to pre-conversion positions and the G-peak FWHM decreased, the D/G ratio of 0.4, up from the starting material at 0.1, and a resistance of about 21kΩ indicated the presence of residual damage was in the graphene lattice, most likely caused by the strain induced by the hydrogenation/dehydrogenation. This strain has its origin in the difference in atomic structure between graphene and graphane as discussed above.

This electrochemical conversion to graphane was also performed on EG grown on n- epitaxial SiC on n+ SiC substrates (Fig. 9). While the substrates are conductive, the presence of a Schottky barrier between SiC and EG [15] ensured that there was sufficient electrical isolation between the substrate and EG layer (Figure 8). This conversion made it possible to take vertical electrical measurements by fabricating EG (graphane)/SiC Schottky diodes on the converted and non-converted areas. Current-voltage and capacitance-voltage measurements were performed on these diodes to extract the key parameters as shown in Table 1. Both EG and graphane Schottky structures exhibited rectification. Most noteworthy are the lower saturation leakage current ($j_o$), lower by a factor of 10 from $10^{-9}$A/cm$^2$ on the graphene to $10^{-10}$ A/cm$^2$ on the graphane, and lower parallel conductance ($g_o$) on the 300µm diodes by a factor of 5.5 at 100kHz, without a decrease in ideality. The ability to perform this process on doped SiC substrates is critical to the application of bandgap engineering of carbon-compounds in advanced electron devices.

**Conclusion**

An electrochemical process was developed to hydrogenate epitaxial graphene grown on SiC substrates to form graphane. This conversion was performed on EG grown on Si- and C-face, SiC substrates, with EG D/G ratio of <0.2. Cyclic voltammetry was used to observe the conversion process and AFM, two-probe resistance measurements and Raman spectroscopy was used to characterize the material before and after hydrogenation. The cyclic votammetry showed both oxidation and reduction peaks, suggesting electrochemical reversibility. The Raman showed an increase in D peak, demonstrating surface functionalization, introduction of a peak at 2930cm$^{-1}$, indicating formation of C-H bonds, and D and 2D peak shifts from pristine EG to the converted graphane, indicating sp$^3$ hybridization. Hydrogenation was also distinguished from lattice damage by thermal reversal back to EG by annealing at 1000°C in vacuum. The Raman spectra of the area after reversal clearly shows disappearance of the C-H peak at ~2930cm$^{-1}$ showing desorption of hydrogen in the material. The D peak shifted back to pre-conversion state and the fluorescence background was no longer present. While the D and 2D peaks shifted back to pre-conversion positions, the D peak's intensity showed that there was some residual damage, D/G ~0.4, most likely caused by the strain of the hydrogenation. Schottky diodes fabricated on graphane formed on doped SiC epilayers showed a significant reduction in current leakage at DC and at 100kHz, adding new possibilities in graphene device engineering.

## Acknowledgements

Kevin M. Daniels acknowledges support from the South East Alliance for Graduate Education and the Professoriate. The authors also acknowledge the Southeastern Center for Electrical Engineering Education (SCEEE) for their support of this work.

Table of Figures

Fig. 1  Schematic of the hydrogenation mechanism of graphene and the resulting material graphane. The hydrogen is attracted and bonds to the pi-bonds of the graphene.

Fig. 2 Diagram of home-built electrochemical setup with current applied though a 10% sulfuric acid solution. Pt wire and exposed graphene are used as the anode and cathode respectively.

Fig. 3.  Cyclic voltammetry curves with a $Hg_2SO_4$ reference (0.67V vs. NHE) showing that the graphane conversion  occurs at ~0.2V below the hydrogen evolution potential in water

Fig. 4 Ivs.t curve for a complete graphane conversion done for an hour. About 1 monolayer of hydrogen was incorporated in 5 monolayers of graphene. This was derived from the total integrated charge

Fig. 5  AFM images showing the surface morphology of (a) Si-face graphene on semi-insulating and (b) C-face graphane on semi-insulating. (c) and (d) shows the surface morphology of Si-face and C-face after graphane conversion respectively. After conversion morphologies showed a slight increase in RMS roughness.

Fig. 6  The  Raman spectra of **graphene** on semi-insulating Si-face SiC**,** the graphene converted to **graphane** and the anneals done for **4 hours** and **50 hours** reversing the conversion.

Fig. 7 Hydrogenation gradient observed with electrochemical conversion of graphene on semi-insulating Si-face SiC

Fig. 8 Schottky diodes fabricated on graphene and graphane on n-  SiC epilayer  grown on the Si-face.

Table I. Key parameters of Schottky diodes on Graphene and Graphane derived from electrical measurements

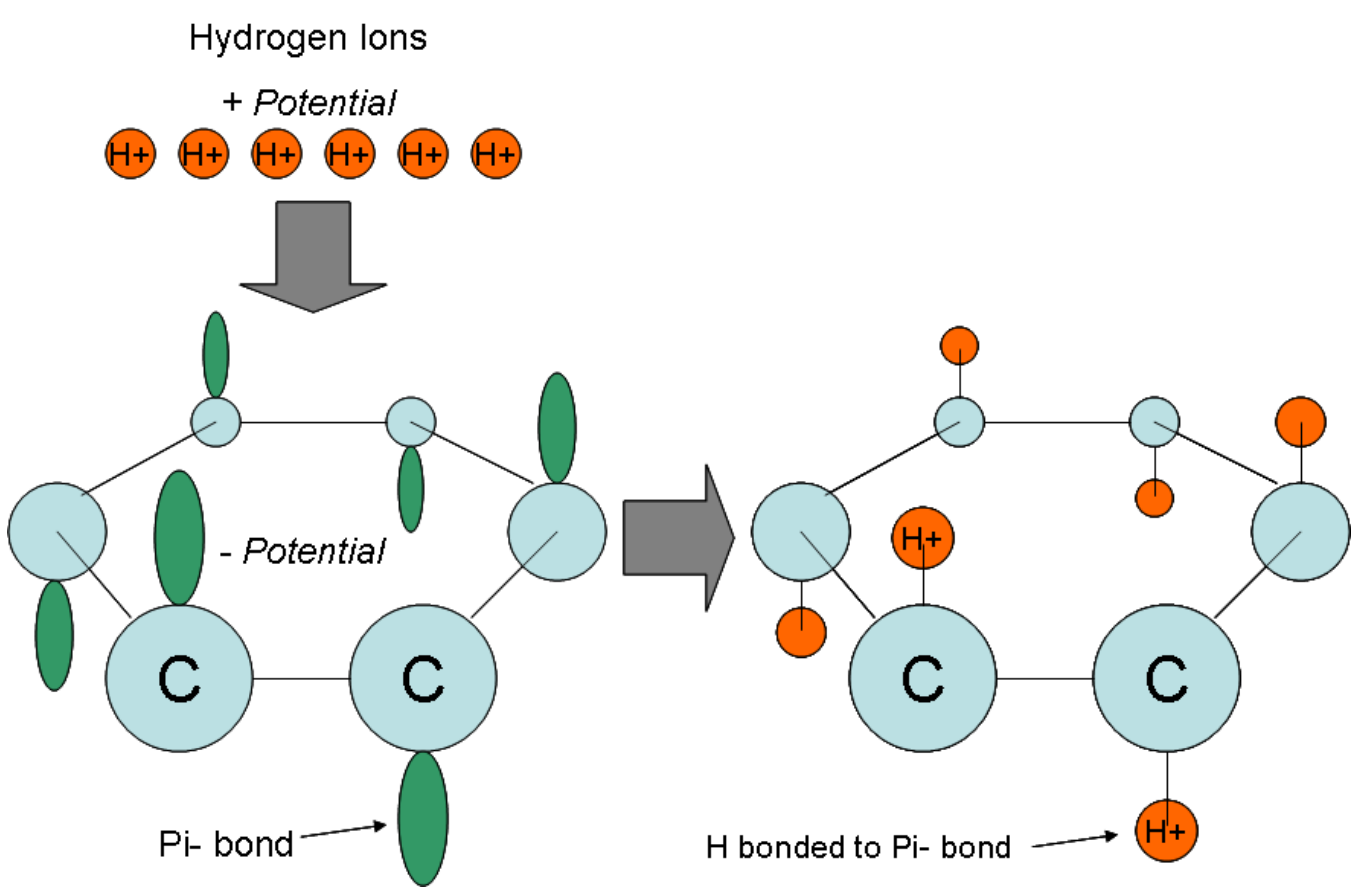

Figure 1

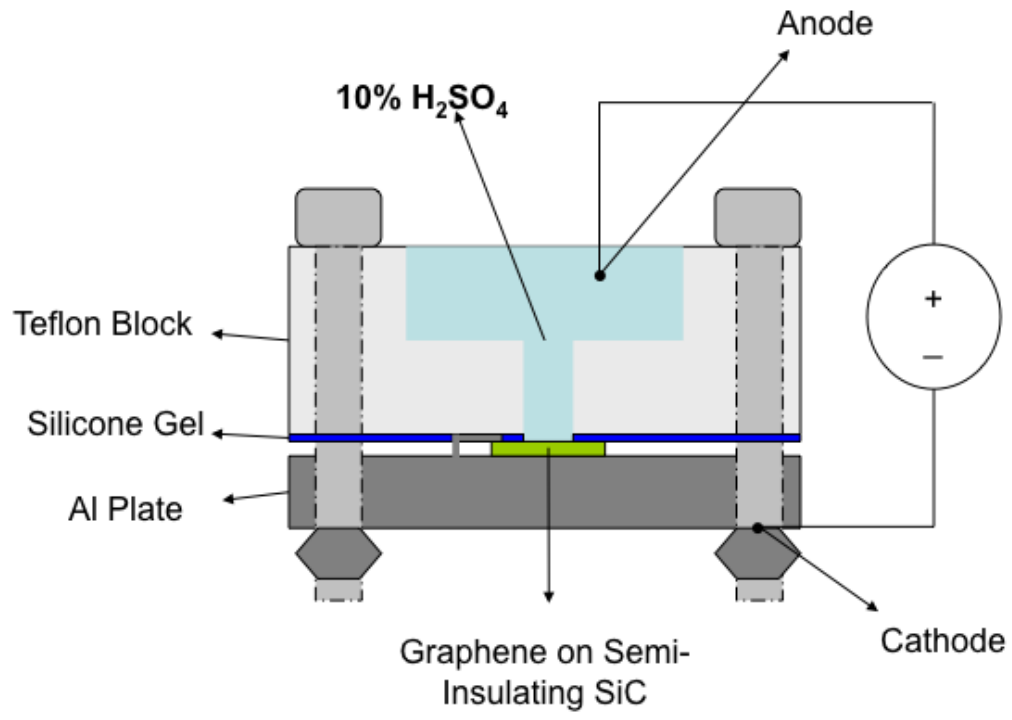

Figure 2

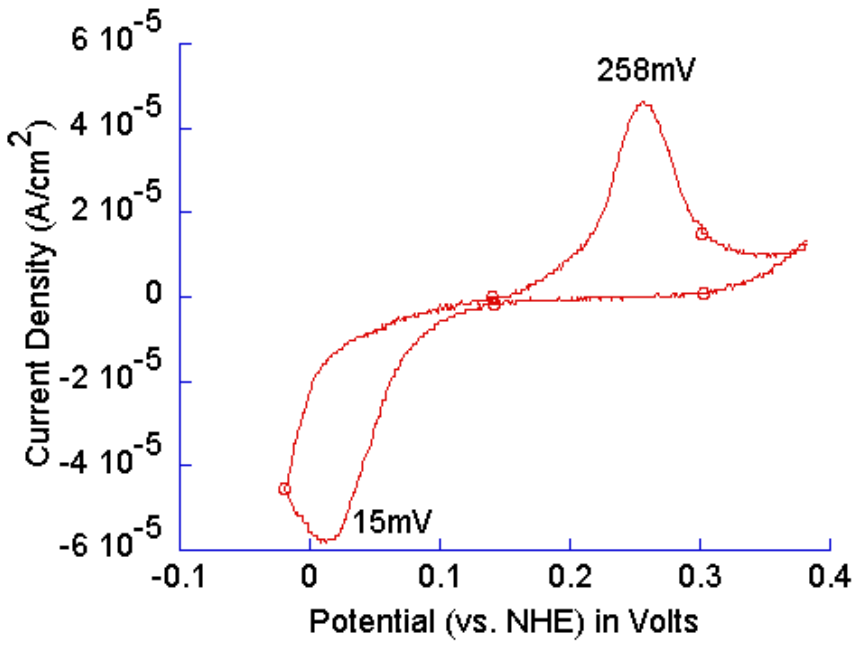

Figure 3

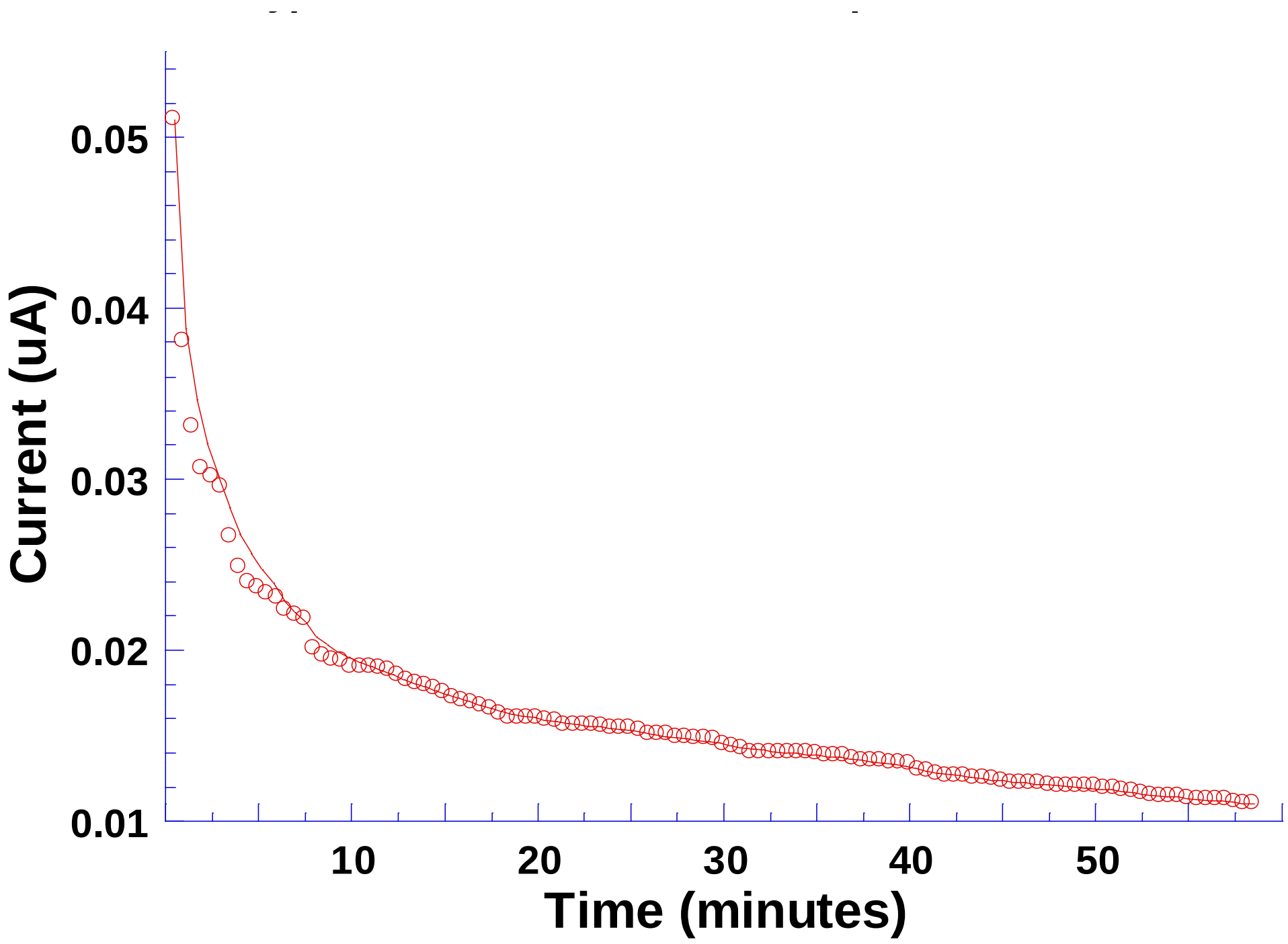

Figure 4

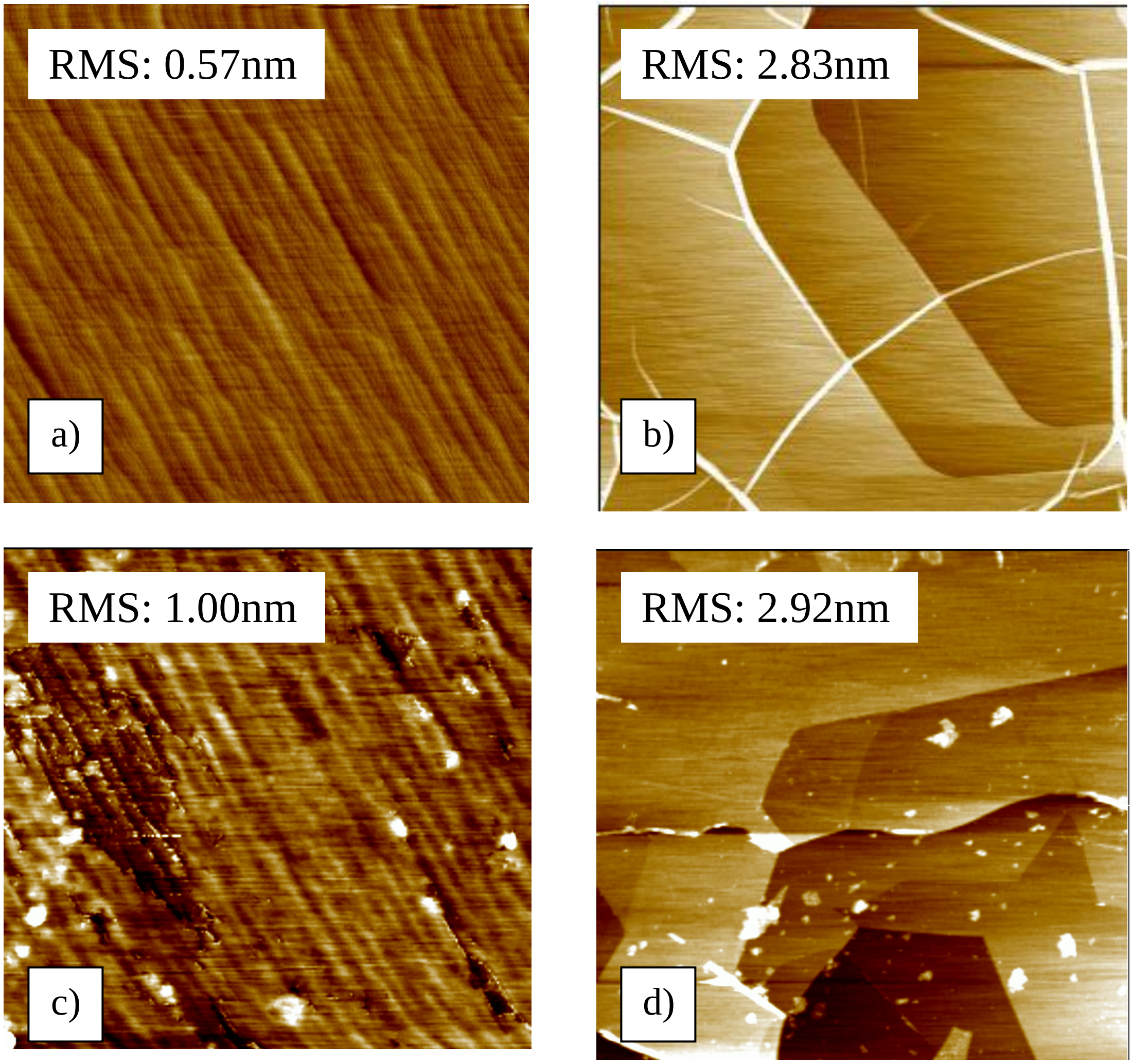

Figure 5

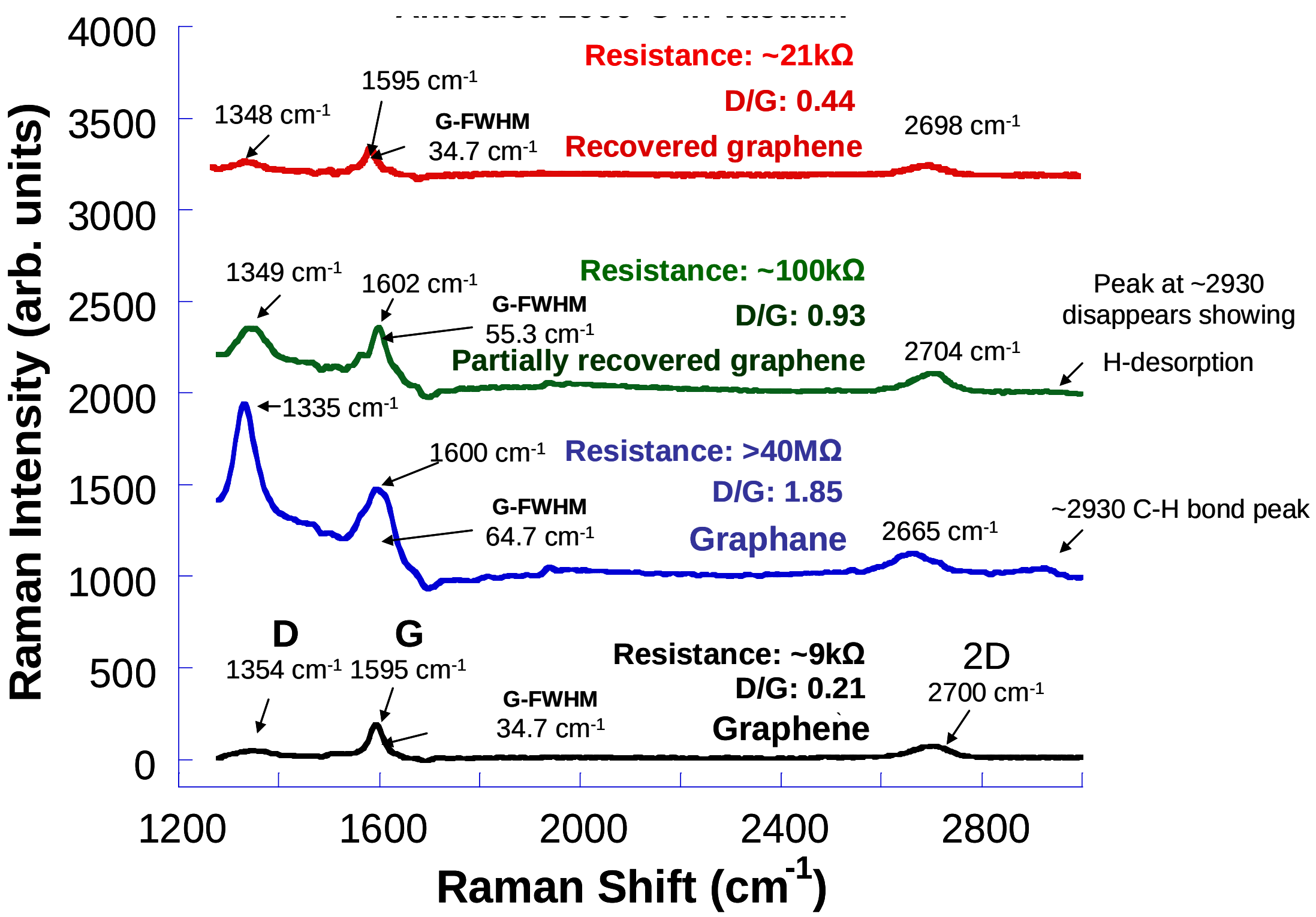

Figure 6

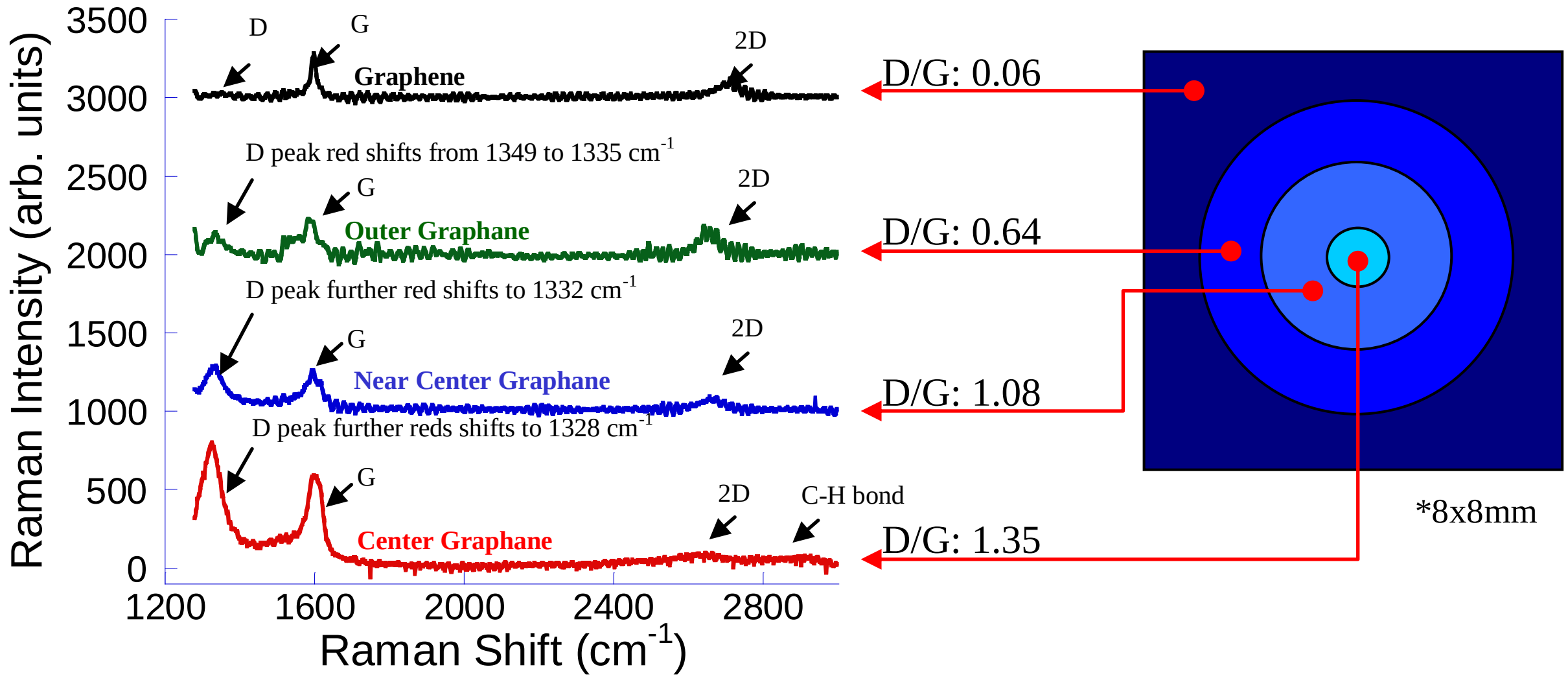

Figure 7

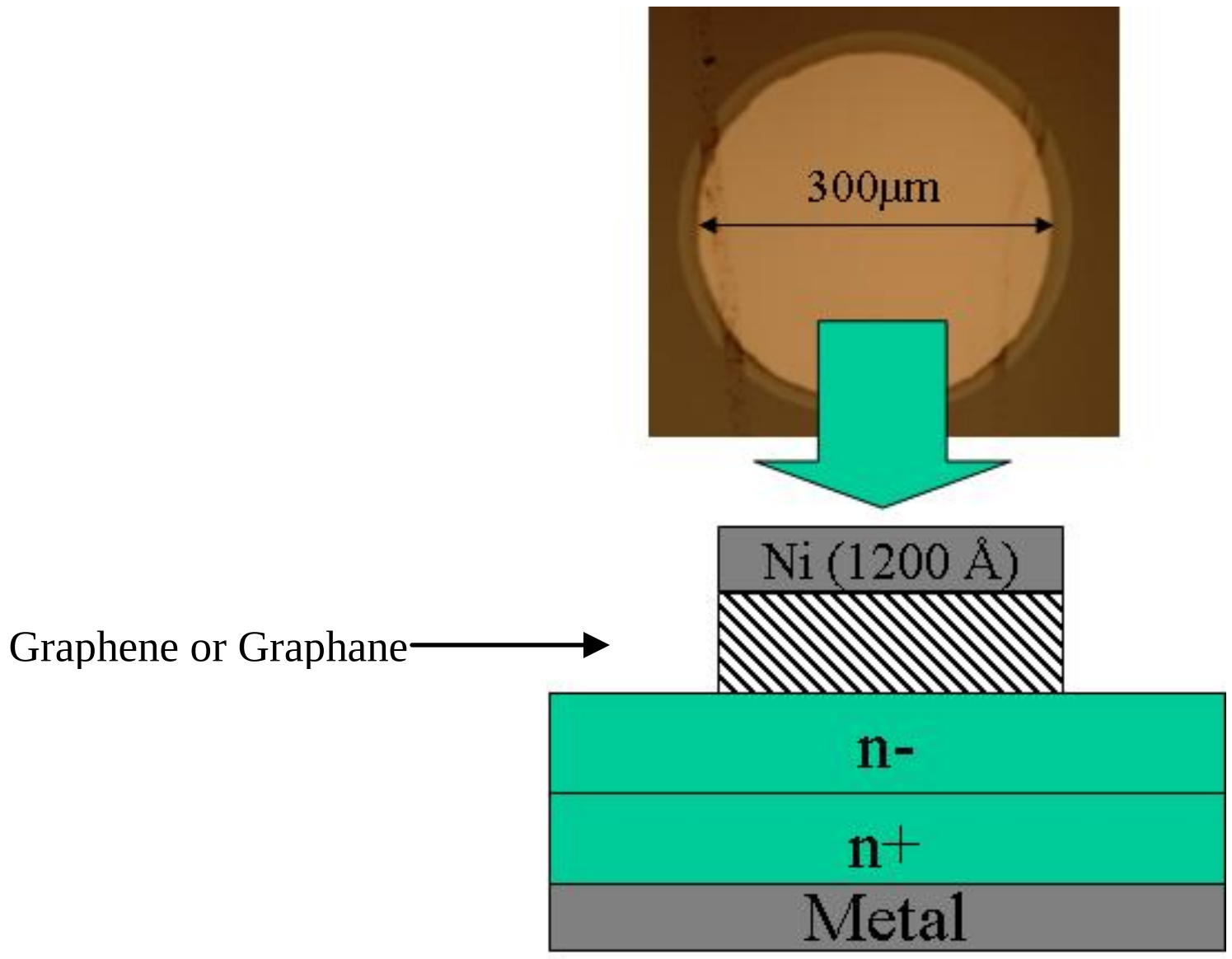

Figure 8

*Table I.*

| | Graphene | Graphane |
|---|---|---|
| $j_o$ (300 µm dia.) **A/cm$^2$** | $10^{-9}$ | $10^{-10}$ |
| **Ideality factor** | 1.21 | 1.11 |
| **Zero bias cap (pF)** | 44.8 | 10.0 |
| **$V_{bi}$ (V)** | 0.527 | 0.631 |
| $g_o$ (300 µm dia.) **S@100kHz** | 9.6x10-7 | 1.72x10-7 |